\documentclass{ws-procs961x669}            
\usepackage{ws-toc,ws-multind}
\usepackage[super,compress]{cite}

\begin{document}

\markboth{M. L\'opez-Corredoira}{Underestimation of H$_0$ error bars}
\wstoc{Underestimation of H$_0$ error bars}{M. L\'opez-Corredoira}

\title{Underestimation of Hubble constant error bars: a historical analysis}

\author{Mart\'\i n L\'opez-Corredoira}

\address{Instituto de Astrof\'\i sica de Canarias, E-38205 La Laguna, 
Tenerife,  Spain\\
and\\
Departamento de Astrof\'\i sica, Universidad de La Laguna, E-38206 La Laguna, 
Tenerife, Spain\\
\email{martin@lopez-corredoira.com}}

\begin{abstract}
An analysis of a historical compilation of Hubble-Lema\^itre 
constant values ($H_0$: 163 data points measured between 1976 and 2019) 
assuming the standard cosmological model gives a $\chi^2$ value of the dispersion
with respect to the weighted average of 580, much larger than the number of points, 
which has  an associated probability that is very low. This means that Hubble tensions
were always present in the literature, due either to the underestimation of 
 statistical error bars associated with the observed parameter measurements, 
or to the fact that  systematic errors were not properly taken 
 into account in many of the measurements.
   
The fact that the underestimation of error bars for $H_0$ is so common might explain 
the apparent 4.4-sigma discrepancy by Riess {\it et al.}
As a matter of fact, more recent precise $H_0$ measurements with JWST data by 
 Freedman {\it et al.} using standard candles in galaxies find there is no tension 
 with CMBR data, possibly indicating that previously Riess {\it et al.} had underestimated their errors. 

Here we have carried out a recalibration of the probabilities with the present
 sample of measurements and we find that x-$\sigma $s deviation is indeed equivalent
 in a normal distribution to the $x_{\rm eq.}\sigma $s deviation, where $x_{\rm eq.}=
 0.83x^{0.62}$. Hence, the tension of 4.4-$\sigma $, estimated between the local Cepheid-supernova 
distance ladder and cosmic microwave background (CMB) data, is indeed a 2.1-$\sigma $
tension in equivalent terms of a normal distribution, with an associated 
probability $P(> x_{\rm eq.})$ = 0.036 (1 in 28). This can be increased to an equivalent 
tension of 2.5-$\sigma $ in the worst cases of claimed 6-$\sigma $ tension, which may in any
 case happen as a random statistical fluctuation. 
 
If Hubble tensions were always present in the literature, and present day tensions are not more
important than previous ones, why, then, is there so much noise and 
commotion surrounding Hubble tension after 2019? It is suggested here that
 this obeys a sociological phenomenon of ``groupthink''.
\end{abstract}

\bodymatter

\section{Introduction}
\label{intro}
{\it [Excerpts from the introductions of Refs. [\citen{Lop22,Wan24}].]}
\\
\\

Few problems in astrophysics have received as much attention in the last years as what is called `Hubble tension'. Hundreds or thousands of papers dedicated to investigating the observations that originate the tension within the standard cosmological model or to propose alternative scenarios have been produced (see other contributions of this parallel session ``Current Status of the $H_0$ and growth tensions: theoretical models and model-independent constraints'' within XVIIth Marcel Grossmann Conference).
The tension was mainly triggered with the claim in 2019 of a Hubble--Lema\^itre constant $H_0$ estimated from the local Cepheid--type Ia supernova (SN Ia) distance ladder being at odds with the value extrapolated from Cosmic Microwave Background (CMB) data, assuming the standard $\Lambda$CDM cosmological model, 
$74.0\pm 1.4$  (Riess et al., Ref. [\citen{Rie19}]) 
and $67.4\pm 0.5$ km s$^{-1}$ Mpc$^{-1}$,\cite{Planck2020} respectively, which gave an incompatibility at the 4.4$\sigma$ level. This tension was later increased up to 6$\sigma $ depending on the datasets considered.\cite{DiV21} 

This tension should not be so surprising, given the number of systematic errors that may arise in the measurements. As a matter of fact, there have always been tensions between different measurements in the values of the Hubble--Lema\^itre constant, which has not received so much attention. Before the 1970s, due to different corrections of errors in the calibration
of standard candles, the parameter continuously decreased its value, making incompatible measurements of different epochs. But even after the 1970s, a tension has always been present. The compilations of values until the beginning of the 2000s showed
an error distribution that was strongly non-Gaussian, with significantly larger probability in the tails of the distribution than predicted by a Gaussian distribution: the 95.4\% confidence-level (CL) limits are 7.0$\sigma $ in terms of the quoted errors.\cite{Che03}

The nature of the possible systematic errors is unknown, and they may dominate over the statistical errors.
Twenty years ago, it was estimated that these systematic errors might be of the order of 
5 km s$^{-1}$ Mpc$^{-1}$ (95\% CL).\cite{Got01} At present, some of these systematic errors may remain without proper correction. As to the possible (non-)universality of the Cepheid period–luminosity relation, 
it was argued that low metallicity Cepheids have flatter slopes, 
and that the derived distance would depend on what relation is used.\cite{Tam03}
We must also bear in mind that the value of $H_0$ is determined without knowing on which scales the radial motion of galaxies and clusters of galaxies relative to us is completely 
dominated by the Hubble--Lema\^itre flow.\cite{Mat95} The homogeneity scale may be much larger than expected, thus giving important net velocity flows on large scales that are incorrectly attributed to cosmological redshifts. Also, values of $H_0$ derived from cosmic microwave background radiation (CMBR) analyses are subject to the errors in the cosmological interpretation of this radiation.\cite{Lop13} That is, CMBR data are interpreted under the assumption of some model, which relates $\Lambda $CDM cosmological model with the power spectrum of CMBR, and there is the possibllity that CMBR can be interpreted with a different model. Moreover, Galactic foregrounds are not perfectly removed,\cite{Axe15,Cre21} 
and these are another source of uncertainties.

Here, we do not contribute a new solution to this Hubble tension in physical terms. Instead we carry out a historical investigation to determine whether or not the given error bars truly represented the dispersion of values in a historical compilation of $H_0$ values. We also show how we can use this knowledge to recalibrate the probabilities of some tension of this kind.

\section{Bibliographical data and statistical analysis}
{\it [Excerpts from Sect. 2 of Ref. [\citen{Lop22}].]}
\\
\\  
  
We use the compililation of 163 values for $H_0$ between the years of 1976 and 2019 compiled by Ref. 
[\citen{Fae20}]. This list was obtained from the NASA Astrophysics Data System
to generate the dataset by carrying out an automated search
of publication abstracts, limiting the search to published papers which include 
values of $H_0$ and their error bars in the paper abstract
itself. Although this selection does not cover the 100\% of the whole literature, it 
gathers most the measurements; this is representative and should not produce any statistical bias. 

The latest measurement we use is the value given by Ref. [\citen{Rie19}], which marked the origin of the present-day Hubble tension. Our goal is investigating the historical records previous to 2019, and not the most recent ones,
because we want to investigate how common the kind of tension pointed out by Ref. [\citen{Rie19}] was until then. 
We wonder whether a 4.4$\sigma $ is something never previously seen or something ordinary. The research of data after Ref. [\citen{Rie19}] in 2019 is not the scope of this paper, as this would belong to an analysis of the post-Hubble tension epoch. Nonetheless, see Ref. [\citen{Wan24}] for an extension of this analysis including the data of 2012-2022.

The~weighted averages of $H_0$ was calculated by weighting each point by the inverse 
variance of that value, and the respective $\chi ^2$ dispersion:
\begin{equation}
\chi ^2=\sum _{i=1}^N\frac{(H_{0,i}-\overline{H_0})^2}{\sigma _i^2} 
.\end{equation} 
We obtain $\overline{H_0}=68.26\pm 0.40$ km s$^{-1}$ Mpc$^{-1}$ and  $\chi^2=575.7$. \citep{Fae20}
For the value of $\chi^2$ calculated using the weighted average, 
the probability that the observed trend is due to chance is $Q = 1.0 \times 10^{-47}$. 
In order to reach a value for $Q$ that is statistically significant ($Q \geq 0.05$), 27 outliers (with more than 2.8$\sigma $ deviation from the average value) must be removed from the data ($n = 136$, $\chi^2=161.3$), producing a value for $Q$ of 0.061. 
Table \ref{Tab:badvalues} lists the 27 values with more than 2.8$\sigma $ deviation.
If instead of 68.26 km s$^{-1}$ Mpc$^{-1}$ we took another value, the $\chi^2$ would be larger, and
the number of points to reject to get $Q \geq 0.05$ would also be larger.

Because of the simplifying assumption made about the neglect of the covariance of each observed measurement, 
the ratio of 27 in 163 is an approximation. A considerable fraction of the published measurements were not independent at all, but consisted of incremental updates to a chain of studies often based on a common foundation of measurements and assumptions.
For a random selection of duplicated points, the reduced $\chi ^2$ is similar, although the ‘effective’ number
of degrees of freedom is lower than the number assuming independence. Probability $Q$ might be larger
when covariance terms are taken into account, but still very low when $Q<<1$.

\begin{table}
{\bf Table 1:} Outliers: measurements of $H_0$ in which $|H_0-\overline{H_0}|>2.8\sigma $, where $\overline{H_0}=68.26$ km s$^{-1}$ Mpc$^{-1}$ is the weighted average of the 163 values of the literature.
\label{Tab:badvalues}
\begin{center}
\begin{tabular}{cccc}
Year & $H_0$ (km s$^{-1}$ Mpc$^{-1}$) & $\frac{|H_0-\overline{H_0}|}{\sigma }$ & Authors  \\ \hline
   1976 &      $50.3\pm 4.3$  &      4.2       &  Sandage \& Tammann \\
   1984 &       $45.0\pm 7.0$ &      3.3       &  J{\~o}eveer  \\
   1990 &       $52.0 \pm 2.0$ &     8.1       &   Sandage \& Tammann \\ 
   1993 &       $47.0 \pm 5.0$ &     4.3       &   Sandage \& Tammann \\ 
   1994 &       $85.0 \pm 5.0$ &     3.3       &   Lu et al. \\ 
   1996 &       $84.0 \pm 4.0$ &     3.9         &  Ford et al. \\        
   1996 &       $57.0 \pm 4.0$ &     2.8        &  Branch et al. \\   
   1996 &       $56.0 \pm 4.0$ &     3.1       &    Sandage et al.\\   
   1998 &       $65.0 \pm 1.0$ &     3.3       &  Watanahe et al. \\  
   1998 &       $44.0 \pm 4.0$ &     6.1     &  Impey et al. \\  
   1999 &       $60.0 \pm  2.0$ &    4.1        &  Saha et al. \\  
   1999 &       $55.0 \pm  3.0$ &    4.4   &  Sandage \\   
   1999 &       $54.0 \pm  5.0$ &    2.9      &  Bridle et al. \\
   1999 &       $42.0 \pm 9.0$ &     2.9    &  Collier et al. \\  
   2000 &       $65.0 \pm 1.0$ &     3.3     &  Wang et al. \\
   2000  &      $52.0\pm  5.5$ &     3.0     &   Burud et al. \\
   2004  &      $78.0\pm  3.0$ &     3.2     &   Wucknitz et al. \\ 
%
   2006  &      $74.9\pm 2.3$  &     3.0      &  Ngeow \& Kanbur \\  
   2006  &      $74.0\pm 2.0$  &    2.9      &   S\'anchez et al. \\
   2008  &      $61.7\pm 1.2$  &  5.7        &  Leith et al. \\
   2012  &      $74.3\pm 2.1$  & 2.9 &  Freedman et al.\\
   2013  &      $76.0\pm 1.9$  &  4.1 &  Fiorentino et al. \\
   2016  &      $73.2\pm 1.7$  &  2.9 &  Riess et al. \\
   2018  &      $73.5\pm 1.7$  &  3.1 &  Riess et al. \\
   2018  &      $73.3\pm 1.7$  &  3.0 &  Follin \& Knox \\
   2018  &      $73.2\pm 1.7$  & 2.9  & Chen et al.  \\
   2019  &      $74.0\pm 1.4$  & 4.1 & Riess et al. \\ \hline
\end{tabular}
\end{center}
\end{table}

The enormous discrepancies between cited uncertainties in derived values of $H_0$ and the much larger dispersions when values from different authors were compared to each other was already realized 
in the 1980s and 1990s.
We now know that there were several underlying reasons for this underestimation of uncertainties in $H_0$ in the measurements before the year $\approx 2000$: errors were usually estimated rather than measured directly via comparison of independent methods; and systematic errors were usually ignored and even the statistical
errors were underestimated. 
Around the turn of the century, when beginning, for example, with the Hubble Space Telescope Key Project on the distance scale, authors began quantifying their uncertainties via multiple independent distance measurements, with separate explicit, and conservative, estimates of the remaining systematic errors, 
e.g. Ref. [\citen{Fre01}]. 
However, looking at the values at Table \ref{Tab:badvalues}, we see many points later than the year 2000 in the list of outliers, which have $>2.8\sigma $ deviation. Certainly, the absolute values of $H_0$ measurement after the year 2000 are closer to the value of the weighted average, because the error bars are also smaller than those ones before 2000. However, the important thing here is not whether the error bars are lower, but whether the error bars are accurately measured, and we have a bunch of values measured after 2000 that are not within this category, assuming that $\overline{H_0}=68.26$ km s$^{-1}$ Mpc$^{-1}$ represents the real value.
If we repeat the whole analysis only
with the subsample of 2001-2019 data (see Sect. 2 of Ref. [\citen{Lop22}]), 
we obtain $\overline{H_0}=69.44\pm 0.26$ km s$^{-1}$ Mpc$^{-1}$ and  $\chi^2=195.2$ for $N=85$ points,
still an unlikely distribution.

We might think that the problem of a too low $Q$ 
can be avoided if we remove some points we suspect to be be wrong, for instance the value
of $61.7\pm 1.2$ km s$^{-1}$ Mpc$^{-1}$ given by Ref. [\citen{Lei08}] (published in a first rank journal of
astrophysics) using SNe Ia. However, removing some values because we do not like them and do not agree the dominant trends is not a valid objective method of data analysis. We can see that there were important Hubble tensions long before the announcement of Riess {\it et al.} and a rigurous statistical analysis cannot decide that some values are more trustable than others.
Ref. [\citen{Lei08}] might have underestimated their errors, and Ref. [\citen{Rie19}] might also have
underestimated the errors. Saying that we trust one team more than another is not a scientific approach.

\section{Recalibration of probabilities}
{\it [Excerpts from Sect. 3 of Ref. [\citen{Lop22}].]}
\\
\\

In Fig. 1 of Ref. [\citen{Lop22}], the frequency of deviation larger than $x\sigma$ 
from the weighted average value $H_{0,{\rm teor.}}=68.26$ km s$^{-1}$ Mpc$^{-1}$ derived
using the whole sample of 163 measurements (including the outliers) is plotted.
Clearly, the probabilities are much higher than those expected in a normal Gaussian error distribution.
For example, in a Gaussian error distribution we should get a $P=2.7\times 10^{-3}$ of obtaining
a deviation higher than 3$\sigma $ (where $\sigma $ is the error of the measurement), but instead we observe that 11.7\% of our measurements get deviations
higher than 3$\sigma $.
The fit of our probability with the frequencies we obtain from the real measurements is:
\begin{equation}
\label{probeq}
P(|H_0-\overline {H_0}| >x\, \sigma )=(0.93\pm 0.06)\times exp[-(0.720\pm 0.013)x]
.\end{equation}
This is equivalent (fit in the range of $x$ between 1 and 12) 
to a number of $\sigma $s deviation in a normal
distribution:
\begin{equation}
x_{\rm eq.}=(0.830\pm 0.004)x^{0.621\pm 0.003}
,\end{equation}
where $x$ is the number of $\sigma $s in the measurement (i.e. $x=\frac{|H_0-\overline{H_0}|}{\sigma }$).
Hence, for instance, a datum that is 3.0$\sigma $ away for the expected value should not be
interpreted as a 3.0$\sigma $ tension, but a 1.6$\sigma $ one in equivalent terms of a normal distribution ($P(>x_{\rm eq.})=0.11$). Likewise, a tension of 4.4$\sigma $ (as for instance claimed by Ref. [\citen{Rie19}]) 
for a Hubble tension) is indeed a 2.1$\sigma $ tension in equivalent terms of a normal distribution, with an associated $P(>x_{\rm eq.})=0.036$ (1 in 28), which is not large but it can occur as a random statistical fluctuation.
For an even larger limit of the tension, at $6\sigma $, as pointed by 
using some different datasets,\citep{DiV21} we would have a equivalent 2.5$\sigma $
with an associate $P(>x_{\rm eq.})=0.012$ (1 in 83), which is still not amazing.

\section{Discussion on the statistical analysis}
{\it [Excerpts from Sect. 4 of Ref. [\citen{Lop22}] and new considerations.]}
\\
\\

One may be tempted to claim that the statistics and treatment of uncertainties in published $H_0$ measurements from the last century cannot be applied to the interpretation of the most recent measurements,
because we know the reason why the error bars measured $>20$ years ago were underestimated, whereas we
are totally certain that our present-day measurements of the error bar are accurate.
However, this statement would be a misunderstanding of the present analysis: we are not here discussing the physics and assumptions (as discussed in the introduction) 
behind those measurements, and certainly we do not have any argument to criticize the techniques of recent $H_0$ measurements. Rather, we are doing here a metaanalysis from a
historical point of view. Of course, Riess {\it et al.} think that their error bars are accurate, but can we be sure that within 30-40 years the same security can be maintained? We can also assume that Sandage, Tammann 
and other very high-qualified specialists measuring $H_0$ in the 1970s were also  convinced that their
measurements were correctly carried out, but it was later proven that they have underestimated their errors.
The idea that scientists in the past were not accurate enough in their statistical analyses
and present-day researchers are doing  a perfect job is not justified from an objective historical point of view.

As a matter of fact, very recent analysis by an important expert team led by Freedman {\it et al.}\cite{Fre24} using
precise James Webb Space Telescope have claimed measurements of $H_0$= 69.85$\pm$1.75(stat)$\pm$1.54(sys) for the Tip 
of the Red Giant Branch stars, 67.96$\pm $1.85(stat)$\pm$1.90(sys) for the J-Region Asymptotic Giant Branch 
stars, 72.05$\pm $1.86(stat)$\pm $3.10(sys) km/s/Mpc for Cepheids, which are compatible with CMBR data. These new 
measurements simply mean that previous Riess et al. measurements\cite{Rie19} based on HST Cepheid/SNeIa distance ladder had understimated their statistical and/or systematic errors. Riess {\it et al.}\cite{Rie24} have expressed their disagreement with an end of the tension, claiming that Ref. [\citen{Fre24}] used a subsample of stars with some 
peculiar deviation with respect to the average in a more complete sample. 
However, it is not clear why the error bars provided by a subsample do not 
overlap with the $H_0$ of the global sample.

History of Science contains plenty of examples of individuals that were pretty sure to be doing a perfect measurement of a parameter, which was later demonstrated to be wrong [e.g., as stated in the famous lecture 
`Cargo Cult Science' by Feynmann\cite{Fey74} about the measurement of the electron charge].
In this sense, this investigation makes sense. The question is: how frequent was to get an underestimation of the error bars in $H_0$ between 1976 and 2019? And we got an answer, which is useful to understand present and future data analyses. 
As scientists, History teaches us many things that we should bear in mind if we do not want to repeat the same errors.

\section{Conclusions on sociology of cosmology}
{\it [Excerpts from Ref. [\citen{Lop23}].]}
\\
\\

In the last five years, no problem in cosmology has received as much attention as what is called “Hubble tension.” Hundreds or perhaps even thousands of papers have investigated the observations that originate the tension within the standard cosmological model or proposed alternative scenarios. 

This Hubble tension is unsurprising, given the number of systematic errors that may arise in the measurements. As a matter of fact, there have always been tensions between different measurements of the values of $H_0$, which have not received much attention. Before the 1970s, due to different corrections of errors in the calibration of standard candles, the value of the parameter continuously decreased, leading to incompatible measurements of different epochs. Even after the 1970s, some tension has always been present. Statistical analyses of the measurements of $H_0$ after the 1970s have also shown that the dispersion of its value is much larger than would be expected in a Gaussian distribution, given the published error bars. The only solutions to understand this dispersion of values are to assume that most of the statistical error bars associated with the observed parameter measurements have been underestimated or to assume that the systematic errors were not properly taken into account.

The fact that the error bars for $H_0$ are so commonly underestimated might explain the apparent discrepancy of values. Indeed, a recalibration of the probabilities with this sample of measurements to make it compatible with a Gaussian distribution of deviations indicates that a tension of 4.4$\sigma $ would be a tension of 2.1$\sigma $ in equivalent terms of a Gaussian distribution of frequencies. Meanwhile, a tension of 6.0$\sigma $ would be a tension of 2.5$\sigma $ in equivalent terms of a Gaussian distribution of frequencies. That is, we should not be surprised to find tensions of 4-6$\sigma $ because they are much more frequent than indicated by the Gaussian statistics, and they stem from underestimation of errors, not from real tensions in the background physics or cosmology.

As said in the introduction, values of $H_0$ derived from CMBR are subject to errors in the cosmological interpretation of CMBR with $\Lambda $CDM, and they are subject to the many anomalies remaining to be solved in CMBR anisotropies. Moreover, Galactic foregrounds in CMBR are not perfectly removed and are an important source of uncertainties. Values of $H_0$ derived from SNIa are affected (and not properly corrected) by dust extinction in SNIa depending on the type of host galaxies, variations of the intrinsic luminosity of SNIa with the age of the host galaxies, etc. Ignoring all these latent variables can only lead to underestimated errors and possible biases. We must also bear in mind that the value of $H_0$ is determined without knowing which scales of the radial motion of galaxies and clusters of galaxies relative to us are completely dominated by the Hubble-Lemaître flow. The homogeneity scale may be much larger than expected, thus giving important net velocity flows on large scales that are incorrectly attributed to cosmological redshifts.

Why, then, is there so much noise and commotion surrounding Hubble tension? There are tens of tensions and problems in the standard model that are more challenging than this,\cite{Lop22a} 
but the cosmological community is currently obsessed with solving the Hubble tension as the key problem in cosmology. 

The answer to this problem has to do with the fact that this tension has been promoted by the dominant groups that control cosmology: the SNIa and CMBR teams, the same ones who promoted the idea of concordance cosmology and dark energy. In the late 1990s, they told astrophysicists something like, ``Now we have a consensus. Everybody should go in this direction,'' and we saw thousands of cosmologists, almost the whole community, moving in this direction like guided herds. Now, one of the supreme leaders, Adam Riess (who won a Nobel Prize in 2011), has said that there is an important problem in cosmology. It is the problem derived from the analysis performed by his own team on SNIa data in comparison with CMBR data, expressing something like, ``There is a tension in cosmology, and everybody should go in this direction.'' Again, we see the cosmologists struggling to understand this tension and reinvent cosmology by adding new patches.

This is called ``groupthink,'' a sociological phenomenon studied in depth, for instance, by psychologist Irving Lester Janis (1918–1990). Orthodox cosmology has an important element of groupthink, of following a leader’s opinion. 
The analysis by Cass R. Sunstein (b. 1954) in his book {\it Conformity}\cite{Sun21} applies to social dynamics in the sciences. Conformity dynamics are particularly pronounced when dealing with very difficult problems. Social experiments in a multitude of contexts clearly show that when a problem is hard to solve, as in cosmology, people tend to follow the crowd, tending to defer to those who are perceived as authorities on the matter. A key mechanism in this collective effect is so-called informational cascades, by which people primarily rely on the signals conveyed by others rather than on independent information. Once this happens, the subsequent statements or actions of few or many others add no new information. They are just following their predecessors. This cascade is very hard to stop, as shown by social networks on the Internet. A reputational cascade develops along with and reinforces an informational one. At this stage, it simply becomes too risky to go against the core consensus.

\

\vfill
\pagebreak

\end{document}